\documentclass[amsmath,amssymb,twocolumn,superscriptaddress]{revtex4-1}

\usepackage{graphicx}
\usepackage{dcolumn}
\usepackage{bm}
\usepackage[detect-all]{siunitx}
\usepackage{hyperref}
\usepackage{xr}

\begin{document}                  

\title{An introduction to classical molecular dynamics simulation for experimental scattering users}

\author{A.~R. McCluskey}
\email{a.r.mccluskey@bath.ac.uk}
\email{andrew.mccluskey@diamond.ac.uk}
\affiliation{Department of Chemistry, University of Bath, Claverton Down,
Bath, BA2 7AY, UK}
\affiliation{Diamond Light Source, Harwell Campus, Didcot, OX11 0DE, UK}

\author{J. Grant}
\affiliation{Computing Services, University of Bath, Claverton Down, Bath, BA2 7AY, UK}

\author{A.~R. Symington}
\affiliation{Department of Chemistry, University of Bath, Claverton Down,
Bath, BA2 7AY, UK}

\author{T.~Snow}
\affiliation{Diamond Light Source, Harwell Campus, Didcot, OX11 0DE, UK}
\affiliation{School of Chemistry, University of Bristol, Bristol, BS8 1TS, UK}

\author{J.~Doutch}
\affiliation{ISIS Facility, Rutherford Appleton Laboratory, STFC, Chilton, Didcot, OX11 0QX, UK}

\author{B.~J. Morgan}
\email{b.j.morgan@bath.ac.uk}
\affiliation{Department of Chemistry, University of Bath, Claverton Down,
Bath, BA2 7AY, UK}

\author{S.~C. Parker}
\affiliation{Department of Chemistry, University of Bath, Claverton Down,
Bath, BA2 7AY, UK}

\author{K.~J. Edler}
\affiliation{Department of Chemistry, University of Bath, Claverton Down,
Bath, BA2 7AY, UK}

\date{\today}

\begin{abstract}
\noindent Classical molecular dynamics simulations are a common component of multi-modal analyses from scattering measurements, such as small-angle scattering and diffraction.
Users of these experimental techniques often have no formal training in the theory and practice of molecular dynamics simulation, leading to the possibility of these simulations being treated as a ``black box'' analysis technique.
In this article, we describe an open educational resource (OER) designed to introduce classical molecular dynamics to users of scattering methods.
This resource is available as a series of interactive web pages, which can be easily accessed by students, and as an open source software repository, which can be freely copied, modified, and redistributed by educators.
The topic covered in this OER includes classical atomistic modelling, parameterising interatomic potentials, molecular dynamics simulations, typical sources of error, and some of the approaches to using simulations in the analysis of scattering data.
\begin{description}
\item[Usage]
Electronic Supplementary Information (ESI) available: All analysis/plotting
scripts and figure files, allowing for a fully reproducible, and automated,
analysis workflow for the work presented is available at
\url{https://github.com/arm61/sim_and_scat_paper} (DOI: 10.5281/zenodo.2556826)
under a CC BY-SA 4.0 license.
\end{description}
\end{abstract}

\maketitle                        

\section{Introduction}

\begin{figure}
\label{fig:growth}
\includegraphics[width=0.48\textwidth]{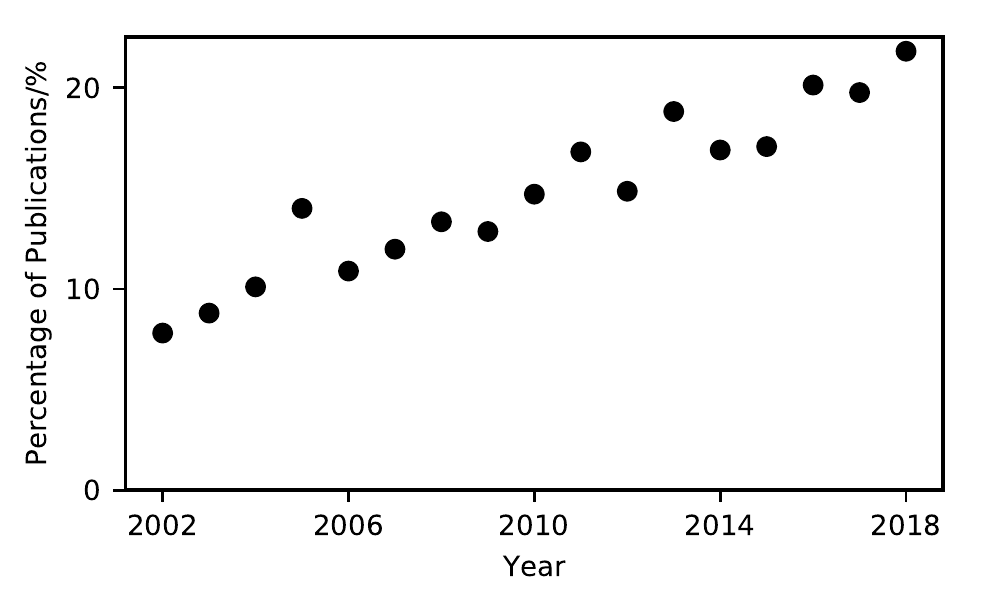}
\caption{The percentage of publications that mention ``small angle scattering'' that also mention ``molecular dynamics'', determined from the numbers of matching Google Scholar results.}
\end{figure}

\noindent The use of molecular dynamics simulations to help analyse and interpret experimental data from small-angle scattering and diffraction studies has grown significantly over the past ten years \cite{pan_molecular_2012,boldon_review_2015,hub_interpreting_2018,ivanovic_temperature-dependent_2018,east_structural_2016,wall_conformational_2014,wall_internal_2018,satoh_multiple_2015}, with the percentage of small-angle scattering publications that mention molecular dynamics reaching more than \SI{20}{\percent} in 2018 (Fig.~\ref{fig:growth}).
Users of scattering and diffraction techniques often have backgrounds in experimental science and may have received little formal training in the theory or practice of computational modelling.
This can lead to molecular dynamics simulations being used as a ``black box'', without understanding the underlying methodologies, or considering possible sources of error.
The use of molecular simulations without any technical understanding can lead to unintended but severe errors.
To help support researchers use of molecular dynamics simulation in their analysis of scattering data while reducing this risk of modelling errors, a number of software tools, such as WAXSiS and SASSIE \cite{chen_validating_2014,knight_waxsis_2015,perkins_atomistic_2016}, have been developed that present easy-to-use, graphical, web-based user interfaces.

A complementary approach is to organise educational activities, such as lectures or workshops, tailored to introduce molecular simulation techniques to audiences of scattering and diffraction users.
One example is the annual ISIS Neutron Training Course, which includes a module titled ``An Introduction to Molecular Dynamics for Neutron Scattering''.
This module covers the fundamentals of classical molecular dynamics simulation, presents applications of these methods in neutron science, and gives students practical hands-on experience with the SASSIE software package \cite{perkins_atomistic_2016}.

While lectures and workshops are an effective tool for education and training, participation can be limited due to difficulties attending in person (due to location or cost) or physical limits on student numbers.
An alternative educational strategy gaining popularity within scientific and engineering communities is the publication of technology-enhanced open educational resources (OERs).
These are courses, lectures, or learning resources published online that are freely available for use by anyone. In addition to their broad accessibility, these resources have permissive ``open'' copyright licences that allow their use in the ``5R activities'': retain, reuse, revise, remix, and redistribute \cite{wiley_open_nodate}.
Publishing a resource as an OER increases its reach and impact, because others may use it in their own teaching not only in its original form, but are free to modify, and redistribute, the material to better suit their aims.
The Jupyter Notebook framework \cite{kluyver_jupyter_2016} has become a popular platform for OERs that teach computational skills, because it allows authors to include instructional text, images, and other media, alongside executable, editable code, in an example of ``literate programming'' \cite{knuth_literate_1984}.
This format encourages students to directly interact with code examples by running, editing, and rerunning these within the source document \cite{barba_cybertraining_2017}, supporting exploratory experiential learning \cite{papert_mindstroms_1993}.

Here, we present an online, open-source, interactive learning resource written to introduce members of the scattering and diffraction community to molecular dynamics simulations.
This OER comprises five lessons that introduce classical molecular dynamics methods and show how these can be used to assist in the analysis of experimental scattering data by the calculation of a simulated scattering profile from the molecular dynamics simulation.
We use the open-source Python library pylj \cite{mccluskey_pylj_2018,mccluskey_arm61/pylj_2019-2} to provide simple, but computationally authentic, examples of simulations, that demonstrate visually and programmatically the conceptual relationships between simulation and scattering techniques.
In this article, we discuss the structure of the resource and describe how a student may get the most from the resource \footnote{In this work \emph{the student} refers to anyone working through the OER, regardless of career position.}.

\section{Assumed prior knowledge}

The OER, entitled \emph{``The Interaction Between Simulation and Scattering''} makes use of the Python programming language to include interactive examples of mathematical and algorithmic content.
To be able to fully use this resource, therefore, requires some knowledge of, or willingness to learn, Python.
We have, however, attempted to design the resource in such a fashion that an in-depth knowledge of Python is not a prerequisite.
It is anticipated that the users of this resource have some experience of at least undergraduate level chemistry or physics, and a commensurate level of mathematical understanding.
This background knowledge will be particularly important for the sections dealing with the functional form and chemical basis of classical interaction potentials.

\section{Resource construction}

The resource is available in two main formats. First, as a series of web pages, hosted at \url{http://pythoninchemistry.org/sim_and_scat}.
Secondly as the source-code repository used to build these webpages, hosted at \url{https://github.com/pythoninchemistry/sim_and_scat} \cite{mccluskey_pythoninchemistry/sim_and_scat_2019}.
The source content consists of a set of Jupyter Notebooks and Markdown files, which are automatically compiled using the \texttt{jupyter-book} tool \cite{lau_jupyter/jupyter-book_2019} to generate the web version.
This system allows the resulting webpages to include text, equations, and figures, which we use to describe key concepts and to explain details of algorithms, as well as Python code blocks, which we use to provide specific examples. The web pages have Thebelab and BinderHub integrations \cite{ragan-kelley_minrk/thebelab_2019, ragan-kelley_jupyterhub/binderhub_2019, jupyter_binder_2018}, which allow students to launch interactive versions of these web pages that allow execution and editing of the included Python code.
We believe that the ability to read the resource as an ``interactive document'' will help students to develop a deeper understanding of the materials than they would from a traditional static document.
The resource is provided under a CC-BY license \cite{creative_commons_creative_2019}, while the \texttt{jupyter-book} software is shared under an MIT license \cite{open_source_mit_2019}, both of which are open and highly permissive.
This allows readers to reuse or remix the material to enhance their own educational platform, and for secondary authors to contribute to improving the source material.

\section{Resource outline}

The resource follows a simple outline that introduces key aspects of molecular dynamics simulations.

\subsection{Home}

The welcome page introduces the resource and gives the student information about how the resource may be used, including details of the Thebelab and BinderHub integrations.
This page also contains details about the permitted use and sharing of the content of the resource, and about the resource license.
This page also includes a list of authors and contributors.

\subsection{Classical methods}

After a brief introduction, this section introduces concepts related to classical simulation methods.
This includes the use of interatomic potential functions, alongside some examples, such as the Lennard-Jones and Buckingham potential models \cite{lennard-jones_determination_1924,buckingham_classical_1938}.
We then introduce the problem of parameterising a potential model, including the use of higher accuracy quantum mechanical calculations to do so.
The presence of off-the-shelf, general potential models are discussed; with the caveat that they may still require system specific optimisation.
Finally, we mention mixing rules; again discussing the possible problems that a user may encounter if applying these blindly to specific systems.

\subsection{Molecular dynamics}

Having introduced the concept of classical interatomic potentials, we then discuss how these are used in molecular dynamics simulations.
We work through how a one-dimensional NVE (constant number, volume, and energy) molecular dynamics simulation may be built, using the Velocity-Verlet algorithm and the Lennard-Jones potential model \cite{swope_computer_1982,lennard-jones_determination_1924}.
The Velocity-Verlet algorithm is introduced in terms of Newton's laws of motion and the generalised equations of motion.
Finally, we discuss a range of key factors that can affect molecular dynamics simulations, including choice of simulation ensembles, setting the range cut-off for an interatomic potential, and the use of periodic boundary conditions.

\subsection{\texttt{pylj} and the interaction with scattering}

The final aspect of the resource covers using molecular dynamics simulations to understand scattering profiles.
This is presented as a practical example, using the open-source \texttt{pylj} package \cite{mccluskey_pylj_2018,mccluskey_arm61/pylj_2019-2}.
We demonstrate a two-dimensional molecular dynamics simulation of argon particles interacting through a Lennard-Jones potential.
The student is first shown a working \texttt{pylj} simulation and invited to interact with the simulation and the custom plotting functionality of \texttt{pylj}.
The concept of a radial distribution function (RDF) is then shown, and the students are given the opportunity to run some \texttt{pylj} simulations with the RDF being output alongside the simulation window.
Next we present the Debye equation \cite{debye_zerstreuung_1915} and show how it may be used to calculate scattering data from a simulation.
The student is invited to observe the effect of simulation temperature on the resulting scattering profile.
We finish by discussing alternative, faster, algorithms for calculating scattering profiles, such as the Fibonacci Sequence or Golden Vectors methods \cite{svergun_solution_1994,watson_rapid_2013}.

\subsection{``Real'' simulation and scattering}

Having shown the development of a scattering profile from an idealised system, we then direct the student to a popular resource for the GROMACS \cite{berendsen_gromacs_1995} molecular dynamics software.
This resource gives a quick introduction to using GROMACS to simulate a lyzosyme molecule in buffer \cite{lemkul_gromacs_nodate}, the student may then use their own simulated trajectory or one that can be downloaded from the OER.
We show how the system may be visualised, introduce the \texttt{MDAnalysis} Python package for the analysis of molecular dynamics trajectories \cite{michaud-agrawal_mdanalysis_2011,gowers_mdanalysis_2016}, and show the scattering profile developed from the lysozyme simulation compared with experimental data \cite{franke_correlation_2015}.
The module finishes by pointing the student to resources to more easily resolve scattering data from molecular simulation, such as SASSIE and CRYSOL \cite{perkins_atomistic_2016,svergun_crysol_1995}.
The focus of this module is to introduce simulation methodologies to users of scattering to aid their understanding, not to derive the exact mechanics of the calculation of scattering from simulation.
Resources for this purpose already exist and have well developed tutorials, so it is not necessary to recreate such software here \cite{perkins_atomistic_2016,svergun_crysol_1995}.

\section{Future outlook}

The aim of this resource is benefit students and researchers interested in combining molecular dynamics analysis with scattering and diffraction experiments.
Our intention in publishing this material as an OER is to promote interest from other parties towards reuse, and remixing the material for their own educational purposes.
We plan to implement this material within training at the ISIS Neutron and Muon Source as well as Diamond Light Source.
Finally, it is hoped, \textit{via} student and community feedback, that the implementation, content, and pedagogical approach of this resource can be increased over time.

\section{Author contributions}

The open education resource was developed, and the manuscript written, by A. R. M. with input from all authors.

\begin{acknowledgements}
A. R. M. is grateful to the University of Bath and Diamond Light Source for co-funding a studentship (Studentship No. STU0149).
B. J. M. acknowledges support from the Royal Society (Grant No. UF130329).
\end{acknowledgements}

\bibliography{bib.bib}

\end{document}